\documentclass[preprint, 11pt, showpacs,preprintnumbers,amsmath,amssymb,nofootinbib]{revtex4}
\usepackage{amsmath, amssymb,graphicx}
\usepackage{epstopdf}
\usepackage {amssymb}
\newcommand{\nc}{\newcommand}
\nc{\ba}{\begin{eqnarray}} \nc{\ea}{\end{eqnarray}}
\newcommand\be{\begin{equation}}
\newcommand\ee{\end{equation}}
\nc{\D}{\overline{\mbox{D3}}}

\nc{\ga}{\gamma} \nc{\tnu}{\tilde{\nu}} \nc{\tmu}{\tilde{\mu}}

\nc{\x}{{\bf{x}}}

\input{epsf.sty}
\begin{document}

\title{Brane Annihilations during Inflation}
\author{Diana Battefeld$^{1)}$}
\email{dbattefe(AT)princeton.edu}
\author{Thorsten Battefeld$^{1)}$}
\email{tbattefe(AT)princeton.edu}
\author{Hassan Firouzjahi$^{2)}$}
\email{firouz(AT)ipm.ir}
\author{Nima Khosravi$^{2,3)}$}
\email{nima(AT)ipm.ir}

\affiliation{1) Princeton University, Department of Physics, NJ
08544, USA} \affiliation{2) School of Physics, Institute for
Research in Fundamental Sciences (IPM), P. O. Box 19395-5531,
Tehran, Iran}\affiliation{3)Department of Physics, Shahid Beheshti
University, G. C., Evin, Tehran 19839, Iran}
\begin{abstract}
We investigate brane inflation driven by two stacks of mobile branes in a throat. The stack closest to the bottom of the throat annihilates first
with antibranes, resulting in particle production and a change of the equation of state parameter $w$. We calculate analytically some observable signatures of the collision; related decays are common in multi-field inflation, providing the motivation for this case study. The discontinuity in $w$ enters the matching conditions relating perturbations in the remaining degree of freedom before and after the collision, affecting the power-spectrum of curvature perturbations. We find an oscillatory modulation of the power-spectrum for
scales within the horizon at the time of the collision,  and a slightly redder spectrum on super-horizon scales. We comment on implications for staggered inflation.
\end{abstract}

\preprint{IPM/P-2010/013}

\maketitle

\tableofcontents


\section{Introduction}
Constructing models of inflation in string theory is an active field, for reviews see e.g.~\cite{HenryTye:2006uv,Cline:2006hu,Burgess:2007pz,McAllister:2007bg,Baumann:2009ni, Mazumdar:2010sa}. 
Because string theory naturally has many dynamical degrees of freedom,
multi-field models of inflation are also common. Examples are inflation from axions \cite{Dimopoulos:2005ac}, tachyons \cite{Piao:2002vf,Majumdar:2003kd}, M5-branes \cite{Becker:2005sg,Ashoorioon:2006wc,Ashoorioon:2008qr} or multiple D-branes \cite{Cline:2005ty, Ashoorioon:2009wa,Ashoorioon:2009sr}
to name a few. 
If many degrees of freedom are present, it is natural for them to decay or stabilize during inflation one after the other. A phenomenological approach to recover some of the effects caused by decaying fields was proposed in \cite{Battefeld:2008py,Battefeld:2008ur,Battefeld:2008qg} and dubbed staggered inflation. \cite{Battefeld:2008py,Battefeld:2008ur,Battefeld:2008qg} relies on coarse graining, assuming that many fields decay in any given Hubble time, which may or may not be the case in concrete models.

In this paper, we set out to investigate a single decay in detail to extract analytically the consequences for cosmological perturbations. 

As a simple and yet non-trivial framework, we choose the KKLMMT \cite{Kachru:2003sx}
brane inflation \cite{dvali-tye,Alexander:2001ks,collection,Dvali:2001fw, Firouzjahi:2003zy,Burgess:2004kv,Buchel,Iizuka:2004ct} set up.
In this proposal, inflation is driven by a brane-antibrane pair, located in a warped throat generated by the background branes and fluxes \cite{Klebanov:2000hb,Giddings:2001yu,Dasgupta:1999ss}.
Inflation ends when the distance between the mobile brane and the antibrane reaches the string scale and a tachyon is formed.  After the brane-antibrane collision, the energy stored in their tensions is released into light closed string modes, which can ignite reheating.  
We extend this proposal by including two mobile stacks with $p_1$ and $p_2$ branes respectively  and $p_1+p_2$ antibranes at the tip of the throat. 
We assume that the stacks are separated from each other so the stack closer to the tip, say stack one, annihilates  first during inflation. This collision results in particle production during inflation and a sudden change in the inflationary potential.  We would like to examine some effects of the produced particles via the change in the equation of state parameter on the power-spectrum of cosmological perturbations.  Inflation ends when the remaining stack of $p_2$ branes annihilates with the background's $p_2$ antibranes. This extension does not improve upon the back-reaction and fine-tuning issues of the KKLMMT proposal \cite{Burgess:2006cb, Baumann:2006th, Baumann:2007ah, Chen:2008au, Cline:2009pu}, but provides a well motivated toy model in string theory.

To handle the analysis, we assume that perturbations in the effective field describing the first stack, $\delta \phi_1$, are taken over by the radiation bath, decaying quickly after the annihilation event. Hence, we focus on perturbations in the remaining field only, $\delta \phi_2$. For this simplification to be consistent, we assume that $p_1 \ll p_2$. We derive asymptotic solutions in the two slow-roll regimes before and after the collision for this field, which we match at the collision time. Treating the mechanism of tachyon formation and the collision as instantaneous events, we use the  Israel junction conditions \cite{Israel:1966rt} to provide the matching conditions, which boil down to the continuity of the Bardeen potential $\Phi$ and the curvature perturbation $\mathcal{R}$ in the case at hand. It is in these matching conditions that the jump in the equation of state parameter enters.  We then compute the power-spectrum of curvature perturbations and find a slightly redder spectrum on scales that were super-horizon at the time of the first stack's annihilation. On sub-horizon scales, we find 
an oscillatory modulation of the power spectrum.

The concrete outline of this paper is as follows: we review and extend the KKLMMT setup in Sec.~\ref{KKLMMT} before deriving the background, slow-roll solution in Sec.~\ref{sec:bgr}. Perturbations are discussed in Sec.~\ref{sec:pertub}, with the matching conditions in Sec.~\ref{Sec:matching} and the Bogoliubov coefficients in Sec.~\ref{sec:bogol}. These coefficients enable the computation of the power-spectrum in Sec.~\ref{sec:powersp}, which we expand for super- and sub-horizon scales in Sec.~\ref{sec:suph} and Sec.~\ref{sec:subh}. We conclude in Sec.~\ref{sec:concl}.


\section{Brane Inflation with Two Stacks
\label{KKLMMT}}
Here we present our set up which is an extension of the KKLMMT proposal of warped brane inflation 
\cite{Kachru:2003sx}.

We start with a warped throat given by the geometry \ba
\label{throat} ds^{2} = h^{-1/2}(r) dx^{\mu} dx^{\nu} + h^{1/2}(r)
(dr^{2} + r^{2} ds_{5}^{2}) \ea where $h(r)$ is the warp factor \ba
h(r)= \frac{L^{4}}{r^{4}} \, . \ea Here $L$ is the Ads scale of the
throat, which is created by $N$ coincident background branes located at $r=0$
\ba \label{L-eq} L^{4}= c\, g_{s} N \alpha'^{2}\,, \ea where $g_{s}$
is the perturbative string coupling, $\sqrt{\alpha'}$ is the string theory length scale and $c$ is a geometric factor
depending on the internal geometry, $ds_{5}^{2}$. For example, for
$S^{5}$ we have $c=4\pi$ while for $T^{(1,1)}$ we get $c=27\,
\pi/4$. Alternatively, one may imagine that the background (\ref{throat}) is created by turning 
on fluxes \cite{Klebanov:2000hb, Giddings:2001yu}. As usual in models of brane inflation, it is assumed that the throat is smoothly glued to the bulk of the compactification. 

\begin{figure}[t]
   \centering
  \includegraphics[width=2in,angle=0]{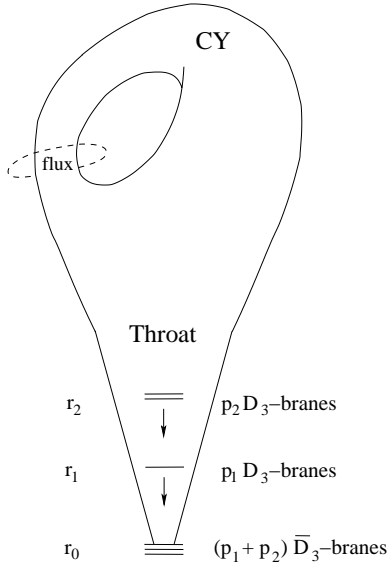}
\caption{A schematic view of the set up. Two stacks of mobile D3-branes are located at $r_{1}$ and $r_{2}$ while $\D$-branes are
located at the bottom of the throat at $r_{0}$. The first
annihilation happens when the physical distance between $r_{1}$ and
$r_{0}$ becomes equal to string scale $l_{s}$, which occurs shortly
after $\eta_{1}=-1$.}
\vspace{0.5cm}
\label{throat1}
\end{figure}
\begin{figure}[t]
   \centering
  \includegraphics[width=4.5in,angle=0]{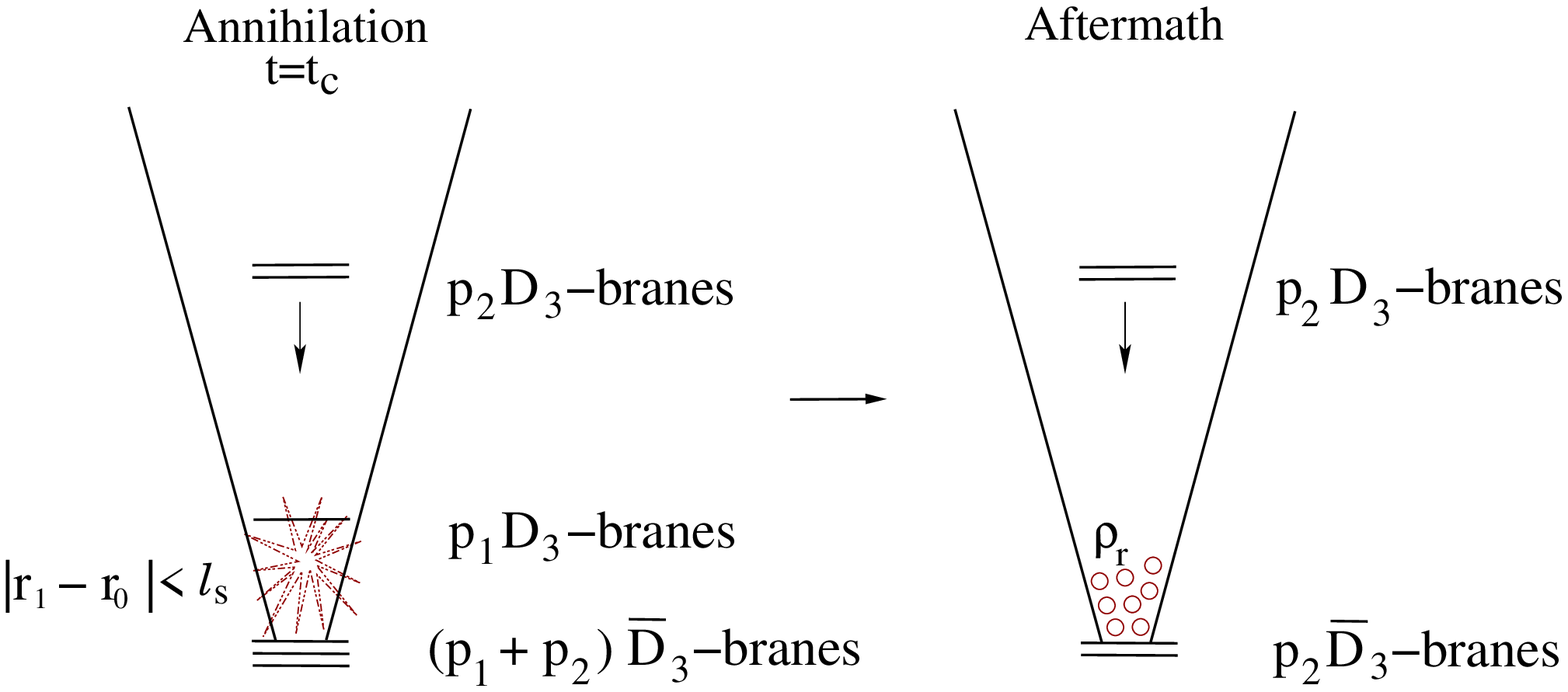}
\caption{A schematic view of the annihilation event at $t_c$. Once the
$p_1$ D3-branes come within $l_s$ to the antibranes, they
annihilate. Their decay products, such as light closed string loops
denoted by $\rho_r$,  cause a jump in the equation of state
parameter $w$ (see Sec.~\ref{Sec:EOSparam}). We require $p_2\gg p_1$ in this study, so that
$\rho_r$ is subdominant after the collision.}
\vspace{0.8cm}
\label{throat2}
\end{figure}

The metric in (\ref{throat}) represents the background geometry, on
top of which we add two stacks of $p_1$ and $p_2$ D3-branes within
the throat at positions $r_{1}$ and $r_{2}$ respectively. At the
bottom of the throat are $p_1+p_2$ $\D $-branes. In accordance to
the KKLMMT proposal, the inflationary period arises from the mutual
attraction of the mobile D3 branes and anti D3-branes plus the
interaction of the $\D $-branes with the background fluxes. We work in
the probe brane approximation, meaning, we assume that the added
stacks of branes and antibranes do not destroy the background. 
For a schematic view of our
set up see {\bf Fig. \ref{throat1}}.

A crucial feature of our model is the annihilation of
brane-antibrane pairs during inflation. To illustrate, during
inflation, the mobile stack of branes closest to the tip of the
throat, comprised of $p_1$ D3-branes, annihilates with $p_1$
$\D$-branes. The effective field corresponding to this stack drops
out of the dynamics, leading to a sudden jump in the potential.
After this annihilation, inflation proceeds driven by the remaining
stack of  branes. Inflation ends when the
second stack annihilates with the $p_2$  antibranes located at the
tip. It is straightforward to generalize this setup to multiple
stacks, $p_I, I>2$, and multiple collisions. However, since our
primary goal is the investigation of a single collision event, we
consider only two stacks. For a schematic of the collision, see {\bf Fig. \ref{throat2}}.

After the annihilation event, the kinetic energies of the branes as
well as potential energies originating from the branes tensions are
transferred into closed string modes \footnote{We ignore other decay channels in this study, such as the productions of cosmic strings \cite{Sarangi:2002yt, Firouzjahi:2005dh}.
Our conclusions primarily depend on a jump of the equation of state parameter; the magnitude of this jump would not change much if other, subdominant decay products different from closed string modes were included. In their presence, we expect only small quantitative changes in i.e. the power-spectrum after the collision.}. In order to simplify the
analysis, we assume that these are in the form of massless modes so
that they behave as radiation. As we shall see in our analysis, it will not change the results if one assumes that the produced particles are in the form of dust. The important effect is the sudden change in the equation of state parameter due to the collision. Furthermore, we assume that the brane
annihilation and tachyon formation is sudden, leading to an
efficient method of converting decay products of the
brane-antibrane annihilation into closed string modes. For a list of phenomenological studies of reheating in models of brane inflation see e.g.  \cite{Shiu:2002xp, Cline:2002it, 
Brodie:2003qv, Barnaby:2004gg, Kofman:2005yz, Chialva:2005zy, Frey:2005jk, Firouzjahi:2005qs, Chen:2006ni, Mazumdar:2008up,Enqvist:2005nc} and the reference therein.

To calculate the inflationary potential, we treat the $p_1
+ p_2\D$-branes as probes in the geometry created by the stacks of
$N$ background branes as well as the additional two stacks of D3-branes \cite{Kachru:2003sx,Baumann:2006th}.  This
leads to the condition $p_{1,2}\ll N$. 
The warp factor in this configuration is then given by \ba
\label{h2} h(r)= \frac{L^{4}}{r^{4}} + \frac{p_1}{N}  \frac{L^{4}}{|
r- r_{1}|^{4}} +\frac{p_2}{N}  \frac{L^{4}}{| r- r_{2}|^{4}} \, .
\ea This form of the metric indicates that the multiple stacks of D3-branes in the absence of antibranes and fluxes
are supersymmetric.

The action of the $ p_1+ p_2 \D$-branes located at the bottom of the
throat, $r_{0}$, is \cite{Kachru:2003sx} 
\ba \label{actionD} S&=&-2
(p_1+ p_2)\, T_{3} \int d^{4}x\,  h(r_{0})^{-1} \, ,
\ea 
where $T_3=
1/(2 \pi)^3 g_s \alpha'^2$ is the D3-brane tension and the factor of
$2$ in Eq. (\ref{actionD}) originates from the combined tension and charge of the
antibranes in this background.

Defining $\phi_{I} \equiv  \sqrt{\, p_I \,  T_{3}} r_{I}$ for
$I=1,2$, $\phi_{A} \equiv  \sqrt{\,(p_1+p_2) T_{3}} r_{0}$ where the
subscript $A$ denotes antibranes, and noting that $r_{0} \ll
r_{1,2}$ \cite{Giddings:2001yu}, we can write the potential for the two scalar fields
$\phi_{I}$ before the collision as \ba \label{potential} V^{-}=
v_{0}^{-} \left[  1- \frac{ b_1}{\phi_{1}^{4}} - \frac{
b_2}{\phi_{2}^{4}}
 \right]  \,,
\ea where \ba v_{0}^{-} \equiv  2 (p_1+p_2)  T_{3} \,
\frac{r_{0}^{4}}{  L^{4}} \quad, \quad   b_I \equiv
\frac{p_I^3\phi_{A}^{4}}{N(p_1+p_2)^2}   \, . \ea

If $r_{1}< r_{2}$ the field $\phi_{1}$ drops out of the dynamics
first. After the annihilation of the first stack, the potential
reduces to \ba \label{V2} V^{+}(\phi_{2}) \simeq   v_{0}^{+} \left[
1 - \frac{ b_2}{\phi_{2}^{4}}
 \right] \quad  \,,
\ea where
\begin{eqnarray}
v_0^{+} \equiv  2  p_2  T_{3}  \frac{r_{0}^{4}}{  L^{4}}\,,
\end{eqnarray}
 leading to a jump in the potential.

After coupling the system to four-dimensional gravity, the total
action is \ba \label{totalS} S=&& \int d^{4} x
\sqrt{- g} \left[ \frac{M_{P}^{2}}{2} R - \frac{1}{2} \partial_{\mu}
\phi_{2} \partial^{\mu} \phi_{2}
   - \left( \frac{1}{2} \partial_{\mu}  \phi_{1} \partial^{\mu} \phi_{1} + V^{-}(\phi_{1}, \phi_{2}) \right)  \theta(-t+t_{c}) \, \right] \nonumber\\
&&+ \int d^{4} x \sqrt{- g} \, \left( { \mathcal{L}_{ra} } -
V^{+}(\phi_{2}) \right) \theta(t-t_{c} ) \ea 
where $M_P$ is the Planck mass related to Newton's constant $G$ via $M_P^2=1/(8 \pi G)$
and  $t_{c}$ indicates
the collision time. In order to respect conservation of
energy at $t_{c}$ we are forced to include an additional
contribution to the action, $\mathcal{L}_{ra}$, corresponding to
radiation created during the annihilation event.

For self-consistency, we require the  initial distance
between branes and antibranes to be greater than the
string scale, $l_{s}= m_{s}^{-1}$, otherwise, the appearance of
tachyons leads to the immediate annihilation of branes and
antibranes. The first inflationary stage terminates at $t=t_{c}$
where the distance between the first stack at $r_{1}$ and the
antibranes at the tip is $l_{s}$. Defining $\phi_{1}(t=t_{c}) =
\phi_{1\, c}$ and using the metric in (\ref{throat}), we obtain
$\phi_{1\,c} = \phi_{A} \exp(l_{s}/L)$.
The onset of tachyon formation and brane annihilation results in the
violation of the slow-roll conditions at the collision time.
Practically, as in \cite{Firouzjahi:2005dh}, the onset of the brane
collision is indicated by $\eta^-_{1}(t_c)=-1$ where $\eta_{1}$,
defined in (\ref{slowrollparameters}), is a slow-roll parameter.


\section{Background: Slow-Roll Inflation \label{sec:bgr}}

In this section, we derive the background dynamics in a homogeneous and isotropic universe where the four dimensional metric is
\ba 
ds^{2} = -dt^{2} + a(t)^{2} d {\vec {\bf
x}}^2\,, 
\ea 
$a(t)$ is the scale factor and $t$ is cosmic
time.  The Friedmann and Klein Gordon equations are 
\ba 
H^{2} =
\frac{\rho}{3M_{P}^{2}}\,, \ea \ba \ddot \phi_{I} + 3 H \dot
\phi_{I} + \frac{\partial V}{\partial{\phi_{I}}} =0\,, 
\ea 
where
$\rho$ is the total energy density containing contributions of
scalar fields and radiation, $\rho_{r}$, after the collision.

The background solutions for $\phi_I$ before the collision are
easily obtained if the system is  in the slow-roll regime, that is
if the slow-roll parameters \ba
 \label{slowrollparameters}
 \epsilon_I \equiv \frac{M_P^2}{2} \left( \frac{V_{,\phi_I }}{V} \right)^2 \quad , \quad
 \eta_{I} \equiv M_P^2 \frac{V_{,\phi_I \phi_I }}{V}  \quad , \quad
 \epsilon \equiv \epsilon_1 + \epsilon_2    \quad , \quad \epsilon_{12}
  = \sqrt{\epsilon_1 \epsilon_2} \, ,
 \ea
 are all much smaller than one. Since there
 is no cross term in the potential, the slow-roll parameter $\eta_{12} \propto V_{\phi_1 \phi_2} $ vanishes. Here, we used the short-hand notation $V_{,\phi_I}\equiv \partial V/\partial \phi_I$. During slow-roll we can approximate
\ba \label{phi-eq} 3 H_{bc} \dot \phi_I + \frac{4 b_I
v_0^-}{\phi_I^5}\simeq 0\,, \ea where we used the potential in
(\ref{potential}) and
\begin{eqnarray}
\label{Hbc}
H_{bc} \equiv \sqrt{\frac{v_0^-}{3 M_P^2}}
\end{eqnarray}
 is the expansion rate in the slow-roll limit before the collision. Eq. (\ref{phi-eq}) can be solved to
\ba \phi_{I\,  in} ^6 - \phi_{I\, c}^6 \simeq \frac{8 b_I v_0^-}{
H_{bc}} t_c\,, \ea where we set the initial time at the start of
inflation to zero and $\phi_I(t=0)\equiv \phi_{I\, in}$  as well as
$\phi_I(t=t_c)\equiv \phi_{I\, c}$. We can replace cosmic time in
terms of the number of e-folds $d N_{e} =  H d t$, so that
 \ba
 \label{phiI-N-eq}
\phi_{I\, in}^{6}- \phi_{I\, c}^{6} \simeq 24 M_{P}^{2}\,  b_I\,
(N_{T} - \Delta N_e) \, . 
\ea 
Here $N_{ T}\simeq 60$ represents the total number of e-folds before the
end of inflation, which needs to be about sixty in order
to solve the flatness and horizon problem, and $\Delta N_e$ is the number of
e-folds from the  time of the first stacks's collision till the end
of inflation, which we assume to be close to $N_T$ throughout this article.
After the first 
stack's collision, the field $\phi_1$ drops out of the dynamics and the
system consists of the field $\phi_2$ and radiation, which is
diluted quickly. Thereafter, $\phi_2$ resumes its slow-roll
evolution, that is the dynamics of $\phi_2$ is given by \ba
\label{phi2-eq} 3 H_{ac} \dot \phi_2 + \frac{4 b_2
v_0^+}{\phi_2^5}\simeq 0\,, \ea where $H_{ac}$ is the expansion rate
in the slow-roll limit after the collision, 
\ba \label{Hac} 
H_{ac} \equiv \sqrt{\frac{ v_0^+}{3 M_{P}^{2}}}  \, . \ea 
Eq.
(\ref{phi2-eq}) is solved by 
\ba 
\label{phi2c} \phi_{2\, c}^{6}-
\phi_{2\, f}^{6} \simeq 24 M_{P}^{2}\,  b_2\, \Delta N_e \, . 
\ea
Here $\phi_{2f}$ is the final value of $\phi_2$ when inflation ends, corresponding to the time when the second stack annihilates with the remaining $p_2$ antibranes at the bottom of the throat. 

The first stack's collision is instigated when $\eta_1^-(t_c)=-1 $, as in  \cite{Firouzjahi:2005dh}, where $\eta_I^\pm$ represents the slow-roll parameters before and after the collision. The $\eta_I$ can be calculated to
 \ba
 \label{eta-}
  \eta_{I}&=& -20 \frac{b_I M_P^2}{\phi_I^6}\nonumber\\
   &=&-\frac{20M_p^2}{NT_3r_0^2}\left(\frac{r_0}{r_I}\right)^6\,.
 \ea
Consequently, the condition $\eta_1^-(t_c)=-1 $ results in
\ba
\label{phi1-c-in}
\phi_{1\, c}^6 = 20 b_1 M_P^2 \quad , \quad
\phi_{1\, in}^6 = 24 b_1 M_P^2 (  N_{T}- \Delta N_e + 5/6 ) \, .
\ea
Similarly, $t_f$, the time of the second stack's collision and the end of inflation, is set by
$\eta_2^-(t_f)=-1 $, which results in $\phi_{2 f}^6 \simeq 20 b_2 M_P^2$.
Combining this with (\ref{phi2c}) and (\ref{phiI-N-eq}) we obtain 
\ba
\label{phi2-f-c}
\phi_{2\, in}^6 \simeq 24 M_P^2 \, b_2 (N_T + 5/6)  \quad , \quad 
\phi_{2\, c}^6 \simeq 24 M_P^2\,  b_2 (\Delta N_{e} + 5/6) \, .
\ea

For later reference it is instructive to express the slow-roll parameters in terms of the number of e-folds $\Delta N_e$ and $N_{T}$.  Using Eqs. (\ref{phi1-c-in}) and 
(\ref{phi2-f-c}) we get
\ba
\label{eta-1-2}
\eta^-_{1\, in} \simeq \frac{-1}{1+ \frac{6}{5}(N_{T} - \Delta N_e)} \quad , \quad
\eta^-_{2\, in} \simeq \frac{-1}{1+ \frac{6}{5} N_T} \quad , \quad
\eta^+_{2\, c} \simeq \frac{-1}{1+ \frac{6}{5} \Delta N_e}\,.
\ea
Here $\eta^-_{1\, in}$ and $\eta^-_{2\, in}$ indicate the slow-roll parameters at the start of inflation whereas $\eta^+_{2\, c}$ indicates the corresponding slow-roll parameter after the first stack's collision. To obtain $\eta^+_{2\, c}$ we assume that radiation is quickly diluted so that the remaining stack resumes its slow-roll motion quickly. 

Regarding the other slow-roll parameters, we note that  $\epsilon_I, \epsilon \ll \eta_{I}$, since
 \ba
 \frac{\epsilon_I}{|\eta_{I }|} \simeq \frac{2 \, p_I}{5 N} \left(\frac{r_0}{r_I} \right)^4
 \ll 1\,,
 \ea
 due to the large value of $N\gg p_I$ and $r_0 < r_I$. This is common in warped brane inflation models such as \cite{Kachru:2003sx}, since the warp factor reduces the height of the potential. This implies that $\epsilon_I \ll |\eta_J|$ which is the main reason why gravitational wave production in these models is suppressed \cite{{Kachru:2003sx}}. 

Since $\phi_1$ drops out of the dynamics after the collision, the
energy associated with the tensions of the $p_1$ antibranes as well
as the kinetic energy of the $p_1$ mobile branes are transferred to
radiation.  During the annihilation event the velocity $\dot
\phi_{2}$ is continuous,  but due to the step in the potential
(\ref{V2}), the acceleration $\ddot \phi_{2}$ makes a jump.  Energy
conservation, $\nabla_{\mu} T^{\mu 0}=0$, dictates the value of
$\rho_{r}$ just after the brane annihilation, 
\ba 
\rho_{r}(t_{c}) =
-{\left[ V  + \frac{1}{2} \dot \phi_{1}^{2} + \frac{1}{2} \dot
\phi_{2}^{2}  \right]}_{\pm} \,, 
\ea 
where we defined $[ f ]_{\pm}
\equiv f(t_{c}^{+}) - f(t_{c}^{-})$. Assuming slow-roll inflation
before the collision, one can neglect the kinetic energy of the
mobile branes compared to the tension of the antibranes and obtains
 \ba
\rho_{r}(t_{c}) \simeq (v_0^- - v_0^+) = 2 p_1 T_{3} \,
r_{0}^{4}/L^{4} \, .
 \ea
 We are interested in the limit where the energy transferred into radiation is small compared to
 the background inflationary potential, that is $\rho_r(t_c)/v_0^- = p_1/(p_1+p_2) \ll 1$ or $p_1\ll p_2$.

Next, we would like to solve for the evolution of the scale factor
after the collision; noting that $\rho_{r}$ scales like radiation
\ba 
\rho_{r}(t) \simeq  (v_0^- - v_0^+)  \left(
\frac{a_{c}}{a(t)} \right)^{4} \,, 
\ea 
where $a_{c}$ is the value of the
scale factor at $t_{c}$, we can approximate the Friedmann equation
for $t>t_c$ by 
\ba 
3 M_p^2 \left( \frac{\dot a}{a} \right)^2 \simeq
v_0^+ +  (v_0^- - v_0^+) \left(\frac{a_c}{a} \right)^4 \,, \ea which
can be solved analytically for $a(t)$, \ba \label{a-biger} a(t) =
\left( \frac{1+ \sqrt {v_0^-/v_0^+}}{2} \right)^{1/2} \left[ 1-
\frac{(v_0^-/v_0^+-1)   e^{-4 H_{ac} (t - t_{c}) } } {  (1+ \sqrt
{v_0^-/v_0^+} )^{2}    } \right]^{1/2} \, a_{c}  \, e^{H_{ac} (t -
t_{c} ) }   \, . 
\ea 
As explained before, $H_{ac}$,  given by Eq.
(\ref{Hac}),
 is the expansion rate after the collision, when radiation is diluted and
the approximate de-Sitter background is recovered from the
interaction of the remaining $p_2$ D3 and $\D$-branes. Comparing
$H_{ac}$ with the expansion rate before the collision, we get
$H_{ac} \simeq H_{bc}  \sqrt{v_0^+/v_0^-}$. Note that $H(t_{c}^{+})
\simeq H_{bc} \neq H_{ac}$ due to the presence of radiation
although $H$ itself is continuous.
Asymptotically, the scale factor in (\ref{a-biger}) approaches \ba
\label{a-biger2} a( t\gg t_c ) \simeq \left( \frac{1+ \sqrt
{v_0^-/v_0^+}}{2} \right)^{1/2} \, a_{c}  \, e^{H_{ac} (t - t_{c} )
}   \, . \ea An interesting implication of (\ref{a-biger2}) is the
increase of the scale factor's amplitude by a factor of $(1+ \sqrt
{v_0^-/v_0^+})^{1/2}/\sqrt{2}$, caused by the transient presence of
radiation.

\section{Perturbations \label{sec:pertub}}
In this section, we consider linear perturbations before and after the collision at $t=t_c$. For $t<t_c$ perturbations are carried by the two
scalar fields (we assume a flat field space metric) and after the collision, by the remaining scalar field and radiation. Perturbations in $\phi_1$ are carried over predominantly by the radiation bath for $t>t_c$ and decay rapidly in the subsequent inflationary phase. Thus, we do not need to follow their evolution.

Our goal in this section is to derive and solve the relevant equations of
motion for perturbations in the surviving field before and after the collision, which need to be matched at
$t_c$.

\subsection{Equations of Motion}
The most general line element including scalar perturbations is
\begin{eqnarray}
ds^2 = -(1 + 2A)dt^2 + 2aB_{,i}dx^i dt + a^2 [(1 - 2\psi)\delta_{ij}
+ 2E_{,ij}]  dx^i dx^j\,,
\end{eqnarray}
where we did not remove any gauge modes yet. The equations of motion
of the perturbations in the scalar fields include couplings to the
metric degrees of freedom and read \cite{Taruya:1997iv, Gordon:2000hv}
\begin{eqnarray}
\ddot{\delta\phi}_I+3H\dot{\delta\phi}_I+\frac{k^2}{a^2}\delta\phi_I+\sum_I
V_{,\phi_I\phi_J}\delta\phi_J=-2V_{,\phi_I}A+\dot{\phi_I}\left[\dot{A}+3\dot{\psi}+\frac{k^2}{a^2}\left(a^2\dot{E}-aB\right)\right]\,,
\end{eqnarray}
with $I=1,2$. One can define gauge invariant metric
perturbations, for instance the two Bardeen potentials
\cite{Mukhanov:1990me}
\begin{eqnarray}
\Phi &=& A+\left(aB-a^2\dot{E}\right)^{.}\,, \label{Bardeen1}\\
\Psi &=&\psi-H \left(aB-a^2\dot{E}\right)\,,\label{Bardeen2}
\end{eqnarray}
that coincide in the absence of anisotropic stress $\Phi=\Psi$. Further, we can define the gauge invariant
Sasaki-Mukhanov variables
\begin{eqnarray}
Q_I=\delta \phi_I+\frac{\dot{\phi}_I}{H}\psi\,.
\end{eqnarray}
These coincide with the field perturbations in the spatially flat
gauge $\psi=0$. In the absence of any other components in the energy
momentum tensor, such as radiation,  and using the perturbed
Einstein equations as well as the background equations of motion one
gets (see \cite{Taruya:1997iv, Gordon:2000hv} for details)
\begin{eqnarray}
\label{Q-eq}
0&=&\ddot{Q}_I+3H\dot{Q}_I+\frac{k^2}{a^2}Q_I+\sum_J\left(V_{,\phi_I\phi_J}-\frac{1}{M_p^2a^3}\left(\frac{a^3}{H}\dot{\phi}_I\dot{\phi}_J\right)^{\!.}\right)Q_J\,.
\end{eqnarray}
Thus, the equations of motion for the Sasaki-Mukhanov variables
decouple from the perturbed Einstein equations for the Bardeen
potentials, and we need not be concerned about the latter. In
deriving (\ref{Q-eq}) we neglected perturbations in the radiation
bath. Our philosophy is to keep the setup as simple as possible,
while retaining some crucial effects onto
perturbations of the annihilation event.

Following \cite{Byrnes:2006fr}, one can show that the equations of
motion simplify during slow-roll to
 \ba
 \label{uEq}
 u_I'' + \left( k^2  - \frac{2}{\tau^2}  \right) u_I \simeq \frac{3}{\tau^2} \sum_J M_{IJ} u_J \, ,
 \ea
 where a prime denotes a derivative with respect to conformal time $\partial/\partial \tau = a\partial/ \partial t$,  $u_I \equiv a Q_I$ and the matrix $M_{IJ}$ is defined as
 \ba
 M_{IJ} \equiv  \left(
\begin{array}{cc}
\epsilon + 2 \epsilon_{1} - \eta_{11} & \quad  2 \epsilon_{12}  \\
2 \epsilon_{12}  & \quad  \epsilon + 2 \epsilon_{2} -
\eta_{22}
\end{array}
\right) \, ,
 \ea
 where we used the slow-roll parameters defined in (\ref{slowrollparameters}) \footnote{We use the potential slow-roll parameters whereas \cite{Byrnes:2006fr} uses the Hubble slow-roll parameters, which coincide at linear order only.}.

\subsection{Before the Collision}
Since $\epsilon_I \ll |\eta_J|$ in warped brane inflation, the matrix 
$ M_{IJ}$ becomes diagonal  and the equations for the $Q_I$ separate,
 \begin{eqnarray}
 u_I^{\prime\prime}+\left(k^2-\frac{\mu_I^2-1/4}{\tau^2}\right)u_I\simeq 0\,,
 \end{eqnarray}
 where 
 \ba
 \mu_I^- \equiv \frac{3}{2} - \eta_I^-
 \ea
 and $\eta_I^-$ is defined in Eq. (\ref{eta-}).

We impose the Minkowski vacuum state $u_I\rightarrow e^{-ik\tau}/\sqrt{2k}$ in the far past ($\tau\rightarrow -\infty$), corresponding to the choice  (see i.e. the review \cite{Bassett:2005xm})
 \ba
 \label{u-}
 u_I^- =  \frac{\sqrt{-\pi \tau}}{2} e^{i \pi (\mu_I^- + 1/2)/2 } H_{\mu_I^-}^{(1)} (- k \tau) \, {\bf e}_I  \, ,
 \ea
 where $H_{\mu_I^-}^{(1)}(x)$ is the Hankel function of the first kind of order $\mu_I^-$ and
 ${\bf e_I}$ are independent unit Gaussian random fields with
 \ba
 < {\bf e}_I > =0 \quad , \quad < {\bf e}_I({\bf k}) \, {\bf e}_J({\bf k'}) >
 = \delta_{IJ} \delta^3 ( {\bf k} - {\bf k'} ) \, .
 \ea
Since perturbations originate during the early stages of inflation well before the collision,  we can use  $\eta_I=\eta^-_{I\, in}$  so that $\mu_I^- = 3/2 - \eta^-_{I\, in}$ from (\ref{eta-1-2}), resulting in 
\ba
\label{mu-12}
\mu_1^- = \frac{3}{2}  + \frac{1}{1+ \frac{6}{5} (N_T - \Delta N_e) }  \quad , \quad 
\mu_2^- = \frac{3}{2}  + \frac{1}{1+ \frac{6}{5}N_T } \,.
\ea

\subsection{After the Collision}
 After the collision, when radiation is diluted away and
  $\phi_2$ resumes its slow-roll motion, the equation for
 $u_2^+$ is given by (\ref{uEq}) with $\eta_{2}^- \rightarrow \eta_{2}^+$ and
 $\mu_2^- \rightarrow \mu_2^+$, which can be integrated to
 \ba
 \label{u+}
 u_2^+ = \frac{\sqrt{-\pi\tau}}{2}\,  e^{i \pi (\mu_2^+ + 1/2)/2 } \, \,
 \left( {\alpha} \,   H_{\mu_2^+}^{(1)} (- k \tau)  +  {\beta}  H_{\mu_2^+}^{(2)} (- k \tau)  \right)  \,.
 \ea
Here $\mu_2^+ \equiv \frac{3}{2} - \eta_{2}^+$, while $\alpha$ and
$\beta$, often referred to as Bogoliubov coefficients, need to be determined by an appropriate matching
procedure near the collision. Using Eq. (\ref{eta-1-2}) we have 
\ba
\label{mu2+}
\mu_2^+ = \frac{3}{2} +  \frac{1}{1+ \frac{6}{5} \Delta N_e } \, .
\ea
We note that $\mu_2^+ \approx \mu_2^-$ since $N_T \sim \Delta N_{e } \gg 1$. 

In deriving the asymptotic solutions in (\ref{u-}) and (\ref{u+}) we treat the slow roll parameters as small and constant. This approximation is common when dealing with inflationary slow-roll models and its validity has been shown numerically. In our model, slow roll is briefly violated during the collision, but due to the short time scale of this collision, we expect this approximation to remain valid.  The time scale of tachyon formation and brane and anti-brane annihilation at the bottom of the throat is given by the inverse of the warped string scale. To trust our four-dimensional effective field theory approximation, the  Hubble expansion rate has to be much smaller than the warped string scale. This indicates that  branes annihilate and slow-roll is violated for less than an e-fold and the slow-roll approximation is restored quickly.

Our goal in the next sections is to calculate $\alpha$ and $\beta$ to compute the power-spectrum. For later use, we define
\ba
\label{alpha-beta}
\alpha \equiv \alpha_1 {\bf e}_1 + \alpha_2 {\bf e}_2  \quad , \quad
\beta \equiv \beta_1 {\bf e}_1 + \beta_2 {\bf e}_2 \,.
\ea

\subsection{Matching Conditions at $t_c$ \label{Sec:matching}}
We treat the collision in a sudden approximation, infusing the
entire energy of $\phi_1$ into the radiation bath $\rho_{r}$ at
$t_c$. We are interested in the limit where the energy transferred into radiation is small compared to the background inflationary potential and $\rho_r(t_c)/v_0^- = p_1/(p_1+p_2) \ll 1$. 

To perform the matching of cosmological perturbations, we demand that both the intrinsic and extrinsic curvatures are continuous on the hyper-surface separating the two inflationary phases \cite{Israel:1966rt, Deruelle:1995kd, Martin:1997zd}. 
As in standard hybrid inflation models \cite{Linde:1993cn, Copeland:1994vg} the 
hyper-surface of the first stack's collision, and thus the transition  between the two phases, is set by a certain value of an inflaton field, $\phi_1 = \phi_c$ for us. Treating the transition as instantaneous, that is assuming $\Delta t\ll H^{-1}(t_c)$,  the cosmological matching conditions relating perturbations before and after the collision become \cite{Zaballa:2009xb,Lyth:2005ze,Zaballa:2006kh}
\begin{eqnarray}
\label{match1}
[ \Phi ]_\pm =0 \quad , \quad 
\label{second-b.c.}[\mathcal{R}]_{\pm}=0\, ,
\end{eqnarray}
where the absence of anisotropic stress was used ($\Phi=\Psi$). These matching conditions are valid for modes with $k\ll (\Delta t)^{-1}$, which includes modes that are within the Hubble horizon at $t_c$. In the above,
the comoving curvature perturbation is defined as
\begin{eqnarray}
\label{defR}
\mathcal{R}&=&\Phi+
\frac{2}{3(1+w)}\left(\frac{\Phi^\prime}{\mathcal{H}}+\Phi\right)
\,.
\end{eqnarray}
The matching conditions in (\ref{match1}) for perturbations are accompanied by the continuity of the scale factor and the Hubble
parameter $\mathcal{H}= a^\prime/a$ (and thus the total energy) at the background level,
\begin{eqnarray}
{[ a ]_\pm=0} \;\;\;\; ,\;\;\;\; {[ \mathcal{H}]_\pm=0} \, ,
\end{eqnarray}
which also follow from the continuity of the induced metric as
well as the extrinsic curvature on the hyper-surface set by the first stack's  collision. 

To make contact with observations, we need to compute the
power-spectrum of the curvature perturbation
$\mathcal{R}$, which is related to the Sasaki-Mukhanov variables via
\cite{Gordon:2000hv}
\begin{eqnarray}
\mathcal{R}&=&\sum_I\left(\frac{\dot{\phi_I}}{\sum_J\dot{\phi_J^2}}\right)Q_I H\,, \nonumber\\
&=& \frac{1}{3 (1+ w) M_P^2 \mathcal{H}}  \sum_I \phi_I^\prime Q_I
\,, \label{defR2}
\end{eqnarray}
if no additional contributions to the energy momentum tensor are present. 

 To impose the matching conditions, we need to know $(\Phi, {\cal R})$ as a function of the
 $Q_I$. Fortunately, ${\cal R}$ is already given in terms of the $Q_I$ in Eq. (\ref{defR2}).
 To obtain $\Phi$ as a function of the $Q_I$, we use the perturbed Einstein equations \cite{Mukhanov:1990me}
 \begin{eqnarray}
 -3\mathcal{H}(\mathcal{H}\Phi+\Psi^\prime)-k^2\Psi&=&\frac{1}{2M_p^2}a^2\delta T_0^{(gi)\,0}\,,\label{PEEQ1}\\
 (\mathcal{H}\Phi+\Psi^\prime)_{,i}&=&\frac{1}{2M_p^2}a^2\delta T_i^{(gi)\,0}\,, \label{PEEQ2}
 \end{eqnarray}
 with the gauge invariant perturbations of the energy momentum tensor
 \begin{eqnarray}
 \delta T_0^{(gi)\,0}&=&\frac{1}{a^2}\sum_I\left(-\phi_I^{\prime 2}\Phi+\phi_I^\prime\delta\phi_I^{(gi)\,\prime}+V_{,\phi_I}a^2\delta\phi_{I}^{(gi)} \right)\,,\\
 \delta T_i^{(gi)\,0}&=&\frac{1}{a^2}\sum_I\phi_I^\prime\delta \phi_{I,i}^{(gi)}\,,
 \end{eqnarray}
 and the gauge invariant field perturbation
 \begin{eqnarray}
 \delta \phi_{I}^{(gi)}=\delta\phi_I+\phi_I^\prime(B-E^\prime)\,.
 \end{eqnarray}
 In the flat gauge we have $\psi=0$ and we can use $Q_I=\delta \phi_I$ as well as  $\delta
\phi_{I}^{(gi)}=Q_I-\phi_I^\prime\Psi/\mathcal{H}$.

  Plugging these into the perturbed Einstein equations and using $\Psi=\Phi$ as well as the background equations of motion, we can solve (\ref{PEEQ2}) for $\Phi^\prime$, insert it into (\ref{PEEQ1}) and arrive after some algebra at
 \begin{eqnarray}
 \label{defPhi}
 \nonumber -k^2\Phi&=&
 \frac{1}{2 M_p^2}\sum_I\left( Q_I^\prime\phi_I^\prime+Q_I\left(a^2V_{,\phi_I}+\frac{3}{2}(1- w) \phi_I^{\prime}\mathcal{H}\right) \right )\ \nonumber\\
 &\simeq& \frac{1}{2 M_p^2}\sum_I \phi_I' Q_I' \, ,
 \end{eqnarray}
 where we used the slow-roll approximation and worked in leading order of the slow-roll parameters, that is we used $1+ w \simeq 0$ and $a^2V_{,\phi_I}+3 \phi_I^{\prime}\mathcal{H} \simeq 0$ before and after the collision.

It is at this point where we make an important approximation to simplify the matching procedure: based on the fact that perturbations in $\phi_1$ are predominantly carried over by perturbations in $\rho_r$, which rapidly redshift, and that $p_2\gg p_1$ so that $\rho_r \ll \rho_{\phi_2}$, we ignore $Q_1$ before the collision just like we ignored $\delta\rho_r$ after the collision. Thus, in order to perform the matching, we approximate (\ref{defR2}) and (\ref{defPhi}) by
\begin{eqnarray}
-k^2\Phi &\approx& \frac{1}{2 M_p^2} \phi_2' Q_2' \,,\label{defR2approx}\\
\mathcal{R}&\approx& \frac{1}{3 (1+ w) M_P^2 \mathcal{H}}  \phi_2^\prime Q_2\,,
\label{defPhiapprox}
\,.
\end{eqnarray}
so that the matching conditions in (\ref{match1}) become
 \ba
 \label{match2}
 \left[  Q_2' \right]_\pm =0 \quad , \quad 
 \left[ \frac{Q_2}{1+w}   \right]_\pm =0 \,,
 \ea
where we used that $\phi_2^\prime$ is continuous. If we now use the asymptotic solutions before and after the collision respectively, (\ref{u-}) and
(\ref{u+}), we can read off the Bogoliubov coefficients $\alpha$ and $\beta$.  Our approximation of neglecting perturbations in radiation and $\delta \phi_1$ is equivalent to
having $\alpha \simeq \alpha_2 {\bf e}_2$ and $\beta \simeq \beta_2 {\bf e}_2$ in Eq. (\ref{alpha-beta}).

Before we proceed, we would like to review our approximations. Our first assumption is that the system is in the slow-roll limit before the collision. A further crucial assumption in our analysis is that  
radiation is quickly diluted after the collision so that the system resumes its slow-roll evolution quickly. In this limit,  (\ref{defR2}) and (\ref{defPhi}) can be used, which are valid for systems composed of scalar fields only.  We note that there have been many phenomenological studies of brane annihilations and reheating in models of brane inflation e.g.~\cite{Barnaby:2004gg, Kofman:2005yz, Chialva:2005zy, Frey:2005jk, Firouzjahi:2005qs, Chen:2006ni}. However, 
we do not know the exact stringy details of tachyon formation and the brane collisions, nor the mechanism of energy transfer into radiation,  especially in a non-trivial background with background fluxes present and collisions between multiple branes and anti-braens, see for example \cite{Jones:2003ae, Jones:2002sia}. Nevertheless, we expect that perturbations in $\phi_1$ are predominantly carried over by perturbations in radiation so that we can use (\ref{defR2approx}) and (\ref{defPhiapprox}) in order to perform the matching.
We still keep some effects of radiation via the change in the equation of state parameter $w$; by means of the Bogoliubov coefficients we can relate the solutions in the two slow-roll inflationary
regimes by using the matching conditions (\ref{match2}). One shortcoming of this treatment is that we do not take into account perturbations in radiation and $\phi_1$, which would enter into the matching conditions. 
Nevertheless, even with this simplified setup, we find interesting effects that we expect to prevail in a more rigorous treatment, which we leave to future study.


\subsection{The Equation of State Parameter \label{Sec:EOSparam}}
Before the collision, the branes move slowly and the slow-roll
conditions are satisfied for both fields so that $w_- \simeq -1$.
Close to the annihilation event, $|\eta^-_{1}|$ becomes of order one;
however, since the annihilation takes place shortly thereafter, the
field $\phi_1$ has no time to speed up and we may use its slow-roll
value, even though slow-roll is violated. Particularly,  we can
neglect its kinetic energy compared to the potential one just before
and after the collision. Similarly, we may use the slow-roll value
for $\phi_2$.

We would like to relate the outgoing solution in the slow-roll approximation in (\ref{u+}) to the incoming solution in (\ref{u-}). We already specified the parameters $\mu_I^-$ and $\mu_2^+$ in (\ref{mu-12}) and (\ref{mu2+}), but we still need to calculate the equation of state parameter for the incoming and outgoing solutions. 

For $t<t_c$ we have
\ba \label{1+w-} 1+ w_-=\frac{\dot{\phi}_1^2+\dot{\phi}_2^2}{3M_p^2H_{bc}^2}  \, ,
 \ea 
 where $H_{bc}$ is the Hubble expansion rate in the slow-roll approximation 
 from (\ref{Hbc}).
After the collision, assuming that radiation is diluted quickly, the slow-roll conditions are satisfied again and the equation of state parameter is given by
\ba
\label{1+w+}
1+ w_+ \simeq  \frac{\dot{\phi}_2^2}{3M_p^2H_{ac}^2} \,,
\ea
so that 
\ba
\label{w-ratio}
\frac{1+ w_+}{1+ w_-}  \simeq \frac{1}{1+ \gamma^2}\, .
\ea
Here we defined 
\ba
\label{gamma-def}
\gamma \equiv \frac{ \dot \phi_1(\tau_c)}{\dot \phi_2(\tau_c)} \simeq  
\sqrt{\frac{p_1}{p_2} } \left( 1+ \frac{6}{5} \Delta N_e \right)^{5/6} \,,
\ea
where we used (\ref{phi1-c-in}) and (\ref{phi2-f-c}).
Similarly, the fractional change in $w$ is
\ba
\label{Deltaw}
\frac{\Delta w}{1+ w_-} \equiv \frac{w_- - w_+}{1+ w_ -} \simeq \frac{\gamma^2}{1+ \gamma^2} \, .
\ea
In the limit $p_1 \rightarrow0$, corresponding to no brane collision during inflation, or $p_2/p_1\gg (\Delta N_e)^{5/3}$, 
we obtain $\gamma \rightarrow 0$, $ w_+ \rightarrow w_-$ and $\Delta w \rightarrow 0$
as expected.

Finally, we note that during slow-roll inflation ${\cal H} \simeq -1/\tau$. In our approximation two slow-roll inflationary regimes are glued to each other with a continuous Hubble expansion rate so there is no jump in conformal time \footnote{To be precise, $\tau_c^-/\tau_c^+\simeq 1+3\gamma^2(1+w_-)/3$, so that the difference between $\tau_c^-$ and $\tau_c^+$ is not only slow-roll suppressed, but also suppressed by $\gamma^2$; as we shall see shortly, $\gamma^2\ll 1$ is needed to guarantee that additional features do not dominate the power-spectrum. }; thus  $\tau$ is continuous on the hyper-surface of the first stack's collision.


\subsection{The Bogoliubov Coefficients \label{sec:bogol}}
At last we are in a position to match the two phases of inflation and calculate the Bogoliubov coefficients 
$\alpha$ and $\beta$, which we need in order to compute the power-spectrum after the first stack's collision. Starting with the matching conditions (\ref{match2}) we obtain
\ba
Q_2'|_- &=& Q_2'|_+ \,,\\
\frac{1+ w_+}{1+ w_-}Q_2  |_- &=& Q_2  |_+\,.
\ea
In the limit where perturbations in radiation and $\delta \phi_1$ are ignored, we have
$\alpha \simeq \alpha_2 \, {\bf e}_2$ and $\beta \simeq \beta_2 \, {\bf e}_2$ in Eq. (\ref{alpha-beta}). With $Q_I = u_I/a$ we find
\ba
\label{alpha-beta-2}
\alpha_2 &=& \frac{-i \pi }{8}e^{i \delta} \left[ \frac{3 \Delta w}{1+ w_-}
 H^{(2)}_{\mu_2^+} H^{(1)}_{\mu_2^-} + 2 x \left( H^{(2)}_{\mu_2^+} H'^{(1)}_{\mu_2^-}  -   \frac{1+w_+}{1+w_-}  H'^{(2)}_{\mu_2^+} H^{(1)}_{\mu_2^-}
\right)  \right]\,, \nonumber\\
\beta_2 &=& \frac{i \pi }{8}e^{i \delta} \left[ \frac{3 \Delta w}{1+ w_-}
 H^{(1)}_{\mu_2^+} H^{(1)}_{\mu_2^-} + 2 x \left( H^{(1)}_{\mu_2^+} H'^{(1)}_{\mu_2^-}  -   \frac{1+w_+}{1+w_-}  H'^{(1)}_{\mu_2^+} H^{(1)}_{\mu_2^-}
\right)  \right]\,,
\ea
 where we defined 
\ba
x&\equiv& -k\tau_c\,,\\
\delta &\equiv& \frac{\pi}{2} (\mu_2^- - \mu_2^+) = \frac{\pi}{2} ( \eta_2^+ - \eta_2^-) \,.
\ea
To derive (\ref{alpha-beta-2}) we used the identity
\begin{eqnarray}
H_{\mu}^{(1)}H_{\mu}^{\prime(2)}-H_{\mu}^{(2)}H_{\mu}^{\prime(1)}=-\frac{4i}{\pi x}\,.
\end{eqnarray}
>From here on, all Hankel functions depend on $x$ and a prime denotes a derivative with respect to $x$. Since $k_c\equiv -1/\tau_c$ is the wave number for which perturbations leave the Hubble horizon at the time of the collision, we see that $x= k/k_c$ smaller or bigger than one discriminates between super- and sub-horizon modes. With the above coefficients and after replacing the Hankel
functions in terms of Bessel functions as well as using the corresponding identity for Bessel-functions,
$J_\mu^\prime(x)Y_\mu(x)-J_\mu(x)Y_\mu^\prime(x)=-2/\pi x$, we obtain 
\begin{eqnarray}
\left<\alpha^*({\bf k^\prime})\alpha({\bf k})\right>-\left<\beta^*({\bf k^\prime})\beta({\bf k})\right> 
&\simeq&\left(|\alpha_2|^2-|\beta_2|^2\right)\delta^3({\bf k}-{\bf k}^\prime)\\
&=& \frac{1+w_+}{1+w_-} \delta^3({\bf k}-{\bf k}^\prime) \\
&\simeq& \frac{1}{1+\gamma^2} \delta^3({\bf k}-{\bf k}^\prime)   \, .
\end{eqnarray}
We note that $|\alpha|^2-|\beta|^2\neq 1$, since the perturbations in $\phi_1$ encoded in
$\alpha_1$ and $\beta_1$ are 
``lost'' at the collision (we assume that they are taken over by radiation and redshifted away rapidly). In the limit $\gamma\rightarrow 0$ we recover the usual normalization of the Bogoliubov coefficients.
However, for arbitrary values of $\gamma $ we checked that by including the contributions of $\alpha_1$ and $\beta_1$ the relation $|\alpha|^2-|\beta|^2 = 1$ holds.


\section{The Power Spectrum \label{sec:powersp}}
We are interested in the power spectrum $\mathcal{P}_{{\cal R}}$ of the
curvature perturbation $\mathcal{R}$, which is defined by (see \cite{Bassett:2005xm} for a review)
\begin{eqnarray}
\delta^3({\bf k}-{\bf k}^\prime)\mathcal{P}_{\mathcal{R}}=\frac{4\pi
k^3}{(2\pi)^3}<\mathcal{R}({\bf k}^\prime)^*\mathcal{R}({\bf k})>
\end{eqnarray}
To relate $\mathcal{R}$ to the
outgoing perturbations after the collision in (\ref{u+}) we use (\ref{defR2}), yielding
\begin{eqnarray}
\mathcal{R}&\simeq & \frac{H}{\dot \phi_2}Q_2\,.
\end{eqnarray}
 At late times
($-k\tau\rightarrow 0$) we can neglect the decaying part of the
Hankel functions in (\ref{u+}), that is
\begin{eqnarray}
H^{(2)}_{\mu_2^+}(-k\tau)\simeq - H^{(1)}_{\mu_2^+}(-k\tau)\simeq
\frac{i}{\pi}\Gamma(\mu_2^+)\left(-\frac{k\tau}{2}\right)^{-\mu_2^+}\,.
\end{eqnarray}
Consequently, we have
\begin{eqnarray}
Q_2\simeq \frac{i}{2 a} \sqrt{- \frac{\tau}{\pi}} \,  e^{i \pi (\mu_2 +
1/2)/2 } \, \Gamma(\mu_2^+)\left(-\frac{k\tau}{2}\right)^{-\mu_2} 
\left(\beta_2 -\alpha_2\right) \, ,
\end{eqnarray}
which results in
\begin{eqnarray}
\mathcal{P}_{\mathcal{R}}&\simeq&
\left( \frac{H^2}{2 \pi \dot \phi_2}   \frac{\Gamma(\mu_2^+)}{\Gamma(3/2)}  \right)^2
\left |\frac{-k \tau}{2} \right|^{3 - 2 \mu_2^+} \left|\beta_2-\alpha_2\right|^2\,.
\end{eqnarray}
The overall amplitude  $\mathcal{P}_{\mathcal{R}} \simeq  (H^2/2\pi \dot{\phi}_2)^2 $
around sixty e-folds before the end of
inflation is set by the COBE normalization $\mathcal{P}_\mathcal{R}\simeq 2 \times 10^{-9}$, which in turn sets the Hubble scale during
inflation.

The important feature in this expression is the transfer function
$\left|\beta_2-\alpha_2\right|^2$, which encodes some effects of the brane annihilation
in terms of the Bogoliubov coefficients. The scalar spectral index, defined as $n_s-1\equiv d \ln
\mathcal{P}_{\mathcal{R}}/ d \ln k$, becomes
\begin{eqnarray}
\label{ns}
n_s-1=n_2-1+\frac{d \ln
\left|\alpha_2-\beta_2\right|^2}{d \ln k}\,,
\end{eqnarray}
where $n_2=4-2\mu_2^+$ is the usual slow-roll result given by
\ba
n_2 \simeq 1- \frac{5}{6 \Delta N_e}\,.
\ea
As in \cite{Kachru:2003sx} and \cite{Firouzjahi:2005dh} one can tune the parameters of the model  ($N$, $g_s, r_0/R$ and $m_s/M_P$) to guarantee a sufficiently long inflationary phase and 
$n_2 \simeq 0.98$.

Compared to standard single field inflationary models, the power-spectrum is altered by the 
transfer function $\left| \alpha_2 - \beta_2  \right|^2$, which encodes some of the effects of the brane collision and particle production 
during inflation.  Additional contributions to the power-spectrum and higher order correlation functions originate via back-scattering of the produced particles onto the inflaton condensate, leading to IR-cascading \cite{Barnaby:2009mc,Barnaby:2009dd}; to estimate this effect, which can dominate the power-spectrum \cite{Barnaby:2009mc} and is not retained in our approach, a better understanding of the brane annihilation process is needed. 

\begin{figure}[t]
   \centering
  \includegraphics[width=4in,angle=0]{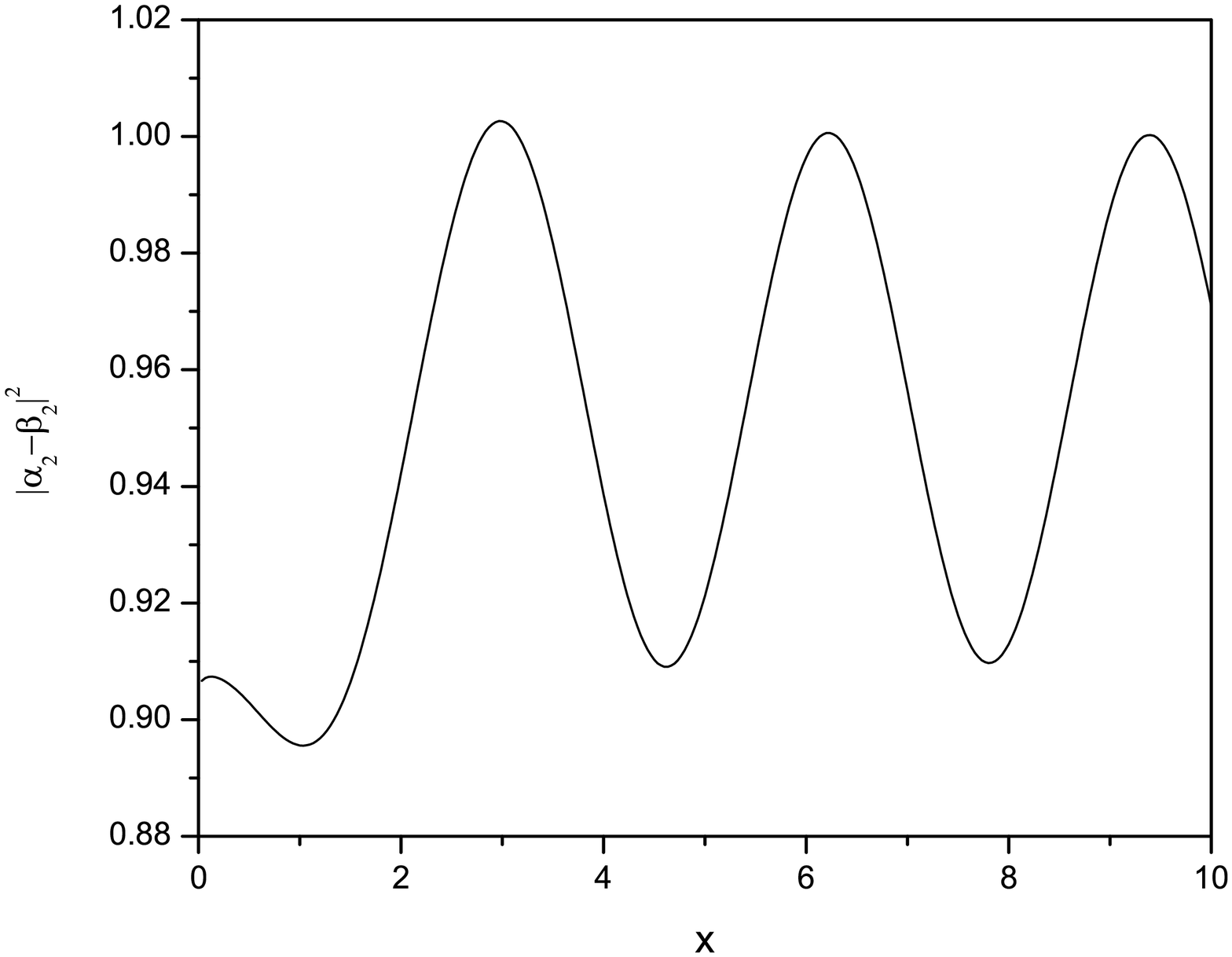}
\caption{The transfer function $|\alpha_2-\beta_2|^2$ from (\ref{alpha-beta-22}) plotted over $x=k/k_c$ for $N_e=60$, $\Delta N_T=58$, $p_1=1$ and $p_2=2.5\times 10^4$. \label{pic:fulltransfer}}
\vspace{0.3cm}
\label{power}
\end{figure}

With $\alpha_2$ and $\beta_2$ from (\ref{alpha-beta-2}), we get
\ba
\label{alpha-beta-22}
| \alpha_2 - \beta_2|^2  &=& \frac{9 \pi^2 }{16}  \left( \frac{\Delta w}{1+ w_-} \right)^2 J_{\mu_2^+}^2  \left(J_{\mu_2^-}^2 + Y_{\mu_2^-}^2  \right) \\
&&+ \frac{3 \pi^2 }{4} \frac{\Delta w}{1+ w_-}
\left[ J_{\mu_2^+}^2 (J_{\mu_2^-}    J'_{\mu_2^-}  + Y_{\mu_2^-}    Y'_{\mu_2^-} ) - 
 \frac{1+w_+}{1+w_-} J_{\mu_2^+} J'_{\mu_2^+}  \left(J_{\mu_2^-}^2 + Y_{\mu_2^-}^2  \right) \right]x \nonumber\\
 &&+
 \frac{\pi^2 }{4} \left[ \left(  J_{\mu_2^+} J'_{\mu_2^-}  -  \frac{1+w_+}{1+w_-} J'_{\mu_2^+} J_{\mu_2^-}   \right)^2 
 + \left(  J_{\mu_2^+} Y'_{\mu_2^-}  -  \frac{1+w_+}{1+w_-} J'_{\mu_2^+} Y_{\mu_2^-}   \right)^2  \right] x^2 \,. \nonumber
\ea
We are interested in the shape of the transfer function in different limits, 
namely  $x\ll 1$,  $x\simeq 1$ and $x\gg 1$, corresponding to modes leaving the horizon before, during and after the collision. The full expression is plotted in {\bf Fig.~\ref{pic:fulltransfer} }, but it is instructive to derive approximate analytic expressions of the transfer function in the limiting cases  $x\ll 1$ and $x\gg 1$.

\subsection{Super-Hubble Scales, $x \ll 1$ \label{sec:suph}}
\begin{figure}[t]
   \centering
  \includegraphics[width=4in,angle=0]{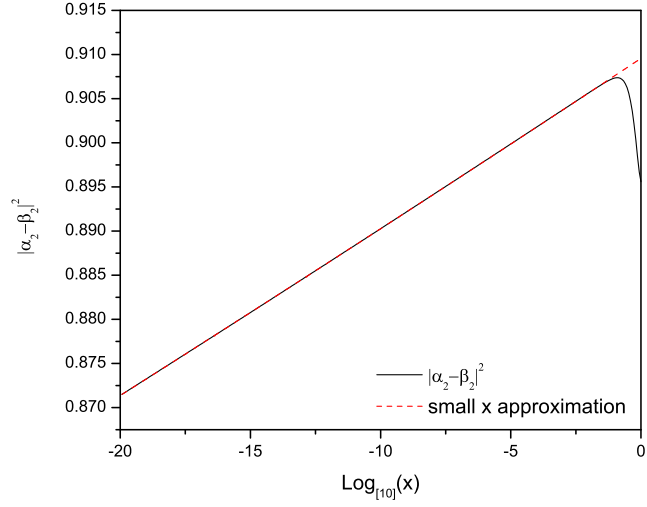}
\caption{The transfer function $|\alpha_2-\beta_2|^2$ in (\ref{alpha-beta-22}), solid line, is compared to the small $x$ limit in (\ref{transfersmallx}), dashed line, for $N_e=60$, $\Delta N_T=58$, $p_1=1$ and $p_2=2.5\times 10^4$. \label{pic:smallxtransfer}}
\label{power}
\vspace{0.5cm}
\end{figure}

The case $x \ll 1$ represents perturbations which cross the Hubble radius long before the collision, $k \ll k_c$. In the absence of entropy perturbations these modes are constant during the two slow-roll phases before and after the collision and we do not expect strong changes as a result of the collision.
To see this, we use the small
$x$ limit of the Bessel functions in (\ref{alpha-beta-22}), which results in
\ba
|\alpha_2 - \beta_2|^2 &\simeq&  \frac{\Gamma^2(\mu_2^-)}{\Gamma^2(\mu_2^+ +1)} 
\left(\frac{x}{2} \right)^{2 (\mu_2^+ -\mu_2^-) }  \left[ \frac{3 \Delta w}{4(1+ w_-)} - \frac{1}{2}  \left(  \mu_2^-+ \mu_2^+\frac{1+ w_+}{1+ w_-}      \right)\right]^2 \label{transfersmallx}\\
&\simeq& \frac{\Gamma^2(\mu_2^-)}{\Gamma^2(\mu_2^+ +1)} 
\left(\frac{x}{2} \right)^{2 (\mu_2^+ -\mu_2^-) }  \left[ \frac{3 \gamma^2}{4(1+ \gamma^2)} - \frac{1}{2}\mu_2^+  \left(  \frac{\mu_2^-}{\mu_2^+}+ \frac{1}{1+ \gamma^2}      \right)\right]^2 \, .
\ea
With $\mu_2^-$ from (\ref{mu-12}) and $\mu_2^+$ from (\ref{mu2+}) we get
\begin{eqnarray}
\mu_2^+ -\mu_2^-&\simeq& \frac{\frac{6}{5}(N_T-\Delta N_e)}{\left(1+\frac{6}{5}\Delta N_e\right)\left(1+\frac{6}{5}N_T\right)}\simeq \frac{5}{6}\frac{N_T-\Delta N_e}{N_T\Delta N_e} \ll 1 \,,\\
\frac{\mu_2^-}{\mu_2^+}&\simeq& \frac{\frac{3}{2}+\frac{1}{1+6N_T/5}}{\frac{3}{2}+\frac{1}{1+6\Delta N_e/5}}\simeq 1-\frac{5}{9}\frac{N_T-\Delta N_e}{N_T\Delta N_e}\approx 1 \, ,
\end{eqnarray}
where $N_T\sim 60$, $1\ll \Delta N_e\lesssim N_T$ is the number of e-folds  
 from the collision until the end of inflation and we expanded in the last step, see {\bf Fig.~\ref{pic:smallxtransfer}} for a plot. We note a mild suppression on super-horizon scales proportional to $x^{2(\mu_2^+-\mu_2^-)}$. This is in agreement with the finding of \cite{Joy:2007na,Joy:2008qd}, whose authors investigated the consequences of a discontinuity in the second derivative of the potential in a single inflaton model, which also caused a jump in $\mu$. The scalar spectral index in (\ref{ns}) picks up an extra contribution on super-horizon scales,
\begin{eqnarray}
n_s\simeq n_2+2\left(\mu_2^+-\mu_2^-\right)\simeq n_2+ \frac{5}{3}\frac{N_T-\Delta N_e}{N_T\Delta N_e}\,,
\end{eqnarray}
which is small compared to the leading order slow-roll contribution. In the absence of a collision, that is for $\mu_2^+=\mu_2^-$ and $\gamma = 0$, we recover  $|\alpha_2 - \beta_2|^2= 1$, as expected.


\subsection{Sub-Hubble Scales, $x \gg 1$ \label{sec:subh}}
\begin{figure}[t]
   \centering
  \includegraphics[width=4in,angle=0]{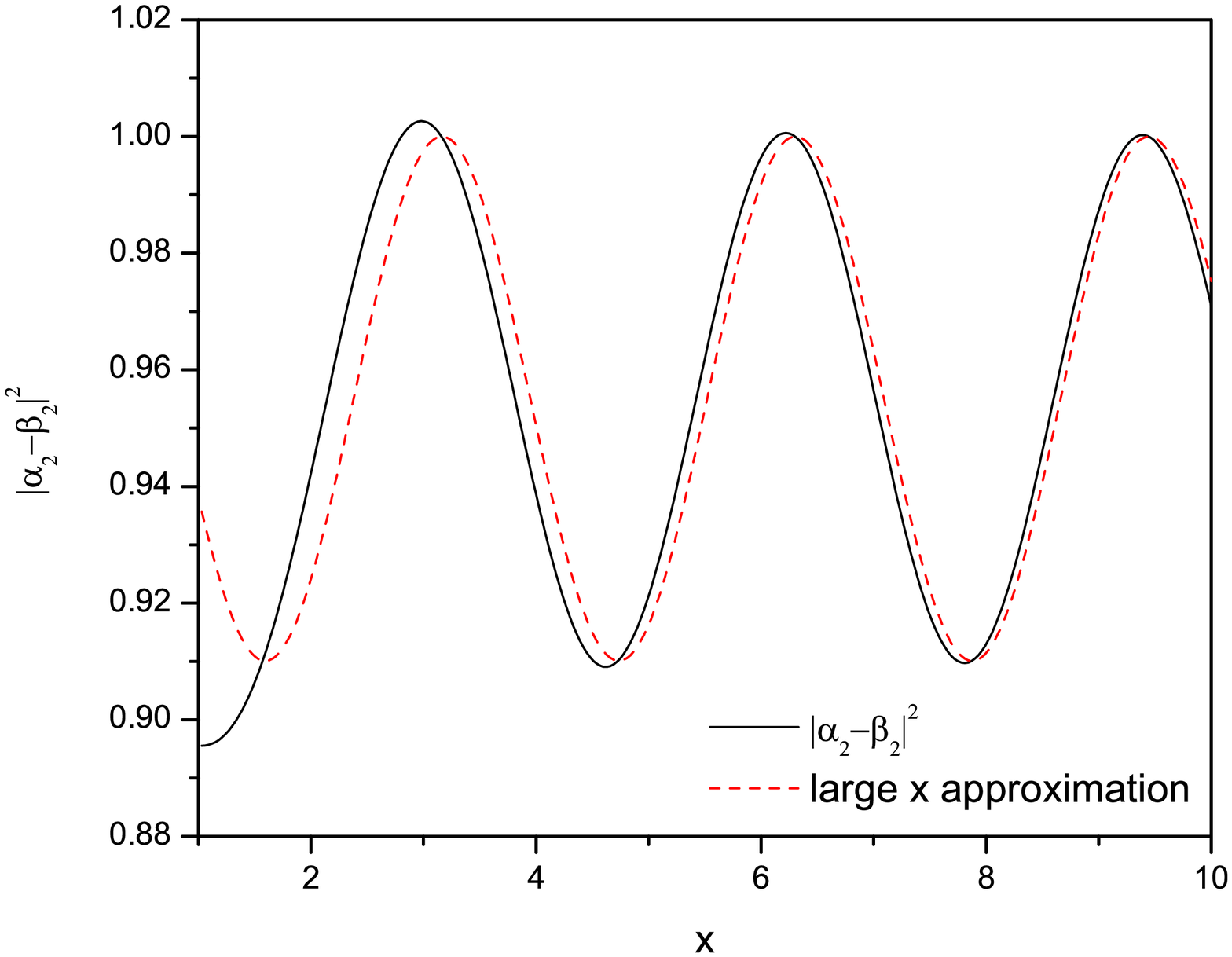}
\caption{The transfer function $|\alpha_2-\beta_2|^2$ in (\ref{alpha-beta-22}), solid line, is compared to the large $x$ limit in (\ref{transferlargex}), dashed line, for $N_e=60$, $\Delta N_T=58$, $p_1=1$ and $p_2=2.5\times 10^4$. \label{pic:largextransfer}}
\vspace{0.5cm}
\label{power}
\end{figure}

In the limit $x \gg1$ we obtain
\ba
|\alpha_2 - \beta_2 |^2 &\simeq& \cos^2\left( x - \frac{\mu_2^+ \pi}{2}  - \frac{\pi}{4} \right)+ \left(\frac{1+w_+}{1+w_-}\right)^2\sin^2 \left( x - \frac{\mu_2^+ \pi}{2}  - \frac{\pi}{4} \right) \label{transferlargex}\\
&\approx& 1-2\gamma^2 \sin^2\left(x - \frac{5 \pi}{12 \Delta N_e}
\right)\,,
\ea
where we expanded for $\gamma \ll 1$. Again, we recover $|\alpha - \beta |^2\rightarrow 1$ in the limit $\gamma\rightarrow 0$. We observe that within the horizon, perturbations are modulated and the power-spectrum picks up an unsuppressed oscillatory mode with amplitude 
\ba 
2\gamma^2\simeq 2 \frac{p_1}{p_2}\left(1+\frac{6}{5}\Delta N_e\right)^{5/3}\,,
\ea 
see {\bf Fig.~\ref{pic:largextransfer}}. If $\Delta N_e\sim N_T\approx 60$, these oscillations are within the observational window of CMB experiments. In order to prevent them from dominating the power-spectrum, we require $2\gamma^2< 0.1$. As a consequence, we need $p_2/p_1> 2.5\times 10^{4}$. On the other hand, the mobile stacks of $p_1$ and $p_2$ branes are assumed to be probe  branes. For consistency, we require large enough background branes (flux charges) corresponding to $N \gtrsim 10^5$. This can be easily obtained in the light of developments in string theory flux compactifications \cite{Giddings:2001yu}, although, the 
large hierarchy $p_2 >  10^{4} p_1 $ may look problematic. 
However, if several branes annihilate successively, we expect these oscillations to be averaged out to some extent, alleviating this bound. 
Thus, it would be interesting to extend the model examined in this paper to include several stacks of branes that annihilate one after the other. Such a setup falls within the framework of staggered inflation, as proposed in \cite{Battefeld:2008py,Battefeld:2008ur}. We plan to investigate such a model numerically in a forthcoming publication \cite{inpreparation}, to bridge the gap between the analytic results of this paper, valid for a single annihilation event, and the analytic results of \cite{Battefeld:2008py,Battefeld:2008qg}, which are expected to hold in the presence of many annihilation events.
 
Small oscillations on top of the power-spectrum, as in {\bf Fig.~\ref{pic:largextransfer}}, have recently caught the attention of the community \cite{Joy:2007na,Joy:2008qd,Romano:2008rr,  Zarei:2008nr, Flauger:2009ab,Biswas:2010si}, since it is possible to mimic outliers in the CMB power-spectrum \cite{Komatsu:2010fb, Ichiki:2009xs, Hamann:2009bz}. 
In \cite{Joy:2007na,Joy:2008qd} the ringing pattern is damped with increasing $x$ whereas in our model, similar to \cite{Romano:2008rr,Flauger:2009ab,Biswas:2010si}, we find no such damping. This indicates that all modes within the horizon are affected by the annihilation event, not only those with $k^{-1}$ close to the horizon scale. However, on the physical grounds one may expect that very small scale modes which effectively live in the Minkowski background should not feel the effect of brane annihilation and particle creation. Our result of Eq. (\ref{transferlargex})  with constant amplitude power spectrum modulation may be due to the sudden  brane annihilation approximation used in our analysis. In practice, the onset of tachyon formation and brane annihilation happens at a time scale given by the inverse of the  warped string scale. This time scale is short enough, i.e. much shorter  than $H^{-1}$, so our approximation for super-horizon perturbations  is valid. However, for wavelengths shorter than the inverse of the warped string mass scale the transition should be smoothed out. As a result, this can remove the power spectrum modulation for wavelength shorter than the inverse warped string scale.

A word of caution regarding the scalar spectral index in (\ref{ns}) might be in order: if one inserts the transfer function (\ref{alpha-beta-22}) into (\ref{ns}) and evaluates $n_s$ for large $x$ one finds a linearly increasing mode -- this mode is an artifact arising from the attempt to describe an oscillatory power-spectrum by means of a simple power law, which is inappropriate.

\subsection{Robustness of Results}
In order to derive the power-spectrum analytically  we have made a series of approximations. 
One important assumption in our analysis  is the neglection of perturbations in the decaying field and radiation, which would both enter in the matching conditions. We plan to relax this approximation in a forthcoming publication -- preliminary results show the same oscillations in the power-spectrum with a comparable amplitude (up to factors of order one), resulting in a similar bound on the number of branes. We ignored other decay channels, for instance the production of cosmic strings at the collision. The oscillatory corrections to the power-spectrum that we find are caused by the jump in the equation of state parameter and the potential. These jumps are present regardless of the exact type of decay products; hence, we expect only minor quantitative changes to the power-spectrum if they were included. We also worked in a sudden decay approximation, matching two asymptotic solutions at the collision-time. In reality, the decay takes some time which smoothes out all discontinuities while slow roll is briefly violated for the decaying field.  As explained previously, the time scale of tachyon formation and brane annihilation is controlled by the inverse of the warped string scale which is much smaller than $H^{-1}$ during inflation in order for our effective four-dimensional approximation to be valid.
In conclusion, our result for super-Hubble perturbations with $x \lesssim 1$ is trusted. However, for wavelengths shorter than the inverse of the warped string scale the transition should be smoothed out. This in turn may remove the unwanted power spectrum modulation for very small scales. 

\section{Discussion and Conclusion \label{sec:concl}}
Motivated by the recent interest in inflationary models with more than one dynamical degree of freedom, we investigated the consequences for cosmological perturbations of a decaying field. As a model we chose one of the best understood models of inflation in string theory, the KKLMMT proposal, and extended it slightly: instead of driving inflation by one brane that annihilates with an antibrane at the end of inflation, we allowed for two stacks of branes with $p_1$ and $p_2$ branes respectively. These two stacks correspond to two inflaton fields at the level of four dimensional effective field theory. If the two stacks are separated initially, the first stack can annihilate with antibranes during inflation, creating closed string loops (radiation) in the process (we ingored other decay channels). We focused on $p_2\gg p_1$, in which case the produced radiation does not interrupt inflation, but causes a slight jump in the equation of state parameter. This setup has all the shortcomings of the original KKLMMT proposal, i.e.~the back reaction of the volume modulus on the inflaton field 
\cite{Burgess:2006cb, Baumann:2006th, Baumann:2007ah, Chen:2008au, Cline:2009pu},
but it provides a well understood framework to investigate the effects of a brane annihilation and particle creation in detail. 

At the perturbed level, we focused on fluctuations of the surviving inflaton field, with the assumption that perturbations in the field corresponding to the first stack of branes are taken over by the produced radiation, which subsequently decay rapidly.  This is our main simplifying assumption. We also ignored contributions from IR-cascading. The asymptotic solutions for fluctuations in the surviving field during the two inflationary regimes, before and after the collision, were matched at the time of the collision according to the Israel junction conditions. It is in this matching procedure that the jump in the equation of state parameter enters. The resulting power-spectrum contains a transfer function, determined by the Bogoliubov coefficients. On scales that are super-horizon during the collision, we find a slightly redder spectrum. On sub-horizon scales, the power-spectrum is modulated by oscillations, or a  ringing, with an amplitude of $\sim 2.5\times 10^3 p_1/p_2$. Such a ringing pattern might help explain glitches in the CMB power-spectrum.  Since these oscillations should not yield corrections bigger than a few percent, we deduce that the remaining stack of branes needs to be comprised of many more branes than the first stack, $p_2\gtrsim  2.5\times 10^4 p_1$. For our probe brane approximation to be valid, we require a large enough background charge, $N \gtrsim 10^5$. It is possible that if several annihilation events were to be superimposed, the amplitude of the oscillations would be reduced and the hierarchy between 
$p_I, I=1,2...$  become milder. Thus, we are naturally lead to the framework of staggered inflation, where many annihilation events are assumed to occur in any given Hubble time. The inclusion of many annihilation events renders a numerical study a necessity, which we plan to come back to in a forthcoming publication.

To summarize, the presence of several dynamical degrees of freedom can lead to additional large contributions to the power-spectrum, such as a ringing pattern, if they start to decay during inflation.

\begin{acknowledgments}

We would like to thank Bruce Bassett, Cliff Burgess, Jim Cline, Misao Sasaki, Ashoke Sen and
Henry Tye for valuable discussions. We also thank Amjad Ashoorioon
and Shahin Sheikh-Jabbari for collaboration at the early stages of
this work and Stephon Alexander, Damien Easson, Anupam Mazumdar, Enrico Pajer, Paul Steinhardt and Brett Underwood for comments.  T.B.~is supported by the Council on Science and
Technology at Princeton University and is thankful for hospitality
at the PI, where this work was instigated. D.B.~is thankful for
hospitality at Princeton University.
\end{acknowledgments}


\end{document}